\documentclass[ two column,rog ]{agutex}

\usepackage {graphics}

\newcommand{\pasj}{{\it Publ. of the Astronom. Soc. of Japan}}
\newcommand{\solphys}{{\it Solar. Phys.}}
\newcommand{\nat}{{\it Nature}}
\renewcommand{\aap}{{\it Astron. Astrophys.}}
\renewcommand{\apj}{{\it Astrophys. J.}}
\renewcommand{\mnras}{{\it Mon. Not. R. Astron. Soc.}}

\authorrunninghead{ROBERT SYCH}
\titlerunninghead{MHD WAVES IN SUNSPOTS}
\input{epsf.sty}
\usepackage {amssymb}

\begin {document}
\title {MHD Waves in Sunspots}

\authors{Robert Sych\altaffilmark{1}}
\altaffiltext{1}{Institute of Solar-Terrestrial Physics SB RAS, p.~o.~box~291, Irkutsk, 664033, Russia;  E-mail: sych@iszf.irk.ru}
 
\begin{article}

\section{Introduction}

Studying magnetohydrodynamic (MHD) waves and oscillations in the solar atmosphere is one of the fastest developing fields in solar physics, and lies in the mainstream of using solar instrumentation data. Wave processes draw our attention due to their possible role in problems of coronal physics, among which: coronal heating, flare energy release and coronal mass ejections, solar wind acceleration. Because the observed properties of waves include the information about propagation medium, they are natural probes providing us with its diagnostics means.

Analyzing quasi-periodic oscillations (QPOs) of the emission above sunspot, first discovered by \citep{1969SoPh....7..351B} and referred as  "umbral flashes" takes a special place in studying wave processes. Although 3-min sunspot oscillations have been observed for decades with sufficient progress in understanding this phenomenon \citep{2006RSPTA.364..313B, 2008sust.book.....T}, we are still far from realizing many problems concerning physics of sunspot wave processes. For example, the relation between 3-min umbral waves and 5-min penumbral traveling waves has not been clear yet. Different studies result in mutually exclusive conclusions: from asserting that there exist prerequisites for penumbral wave emergence from sunspot umbral oscillations \citep{2001A&A...375..617C} to addressing them as independent phenomena \citep{2000A&A...354..305C, 2004A&A...424..671K}. Recent studies \citep{2006A&A...456..689T, 2007A&A...463.1153T,2008sust.book.....T} showed that umbral flares and penumbral traveling waves are a manifestation of a common wave process with wave fronts propagating through the sunspot umbra and penumbra.

It is generally accepted that magnetoacoustic waves move at a local sound speed along magnetic field lines from the photospheric layers through the chromosphere into the corona. However, it is still unclear how high they propagate into the corona. What determines their properties in the corona? How does the sunspot wave cut-off frequency vary? How are the oscillation parameters spatially distributed depending on the oscillation period? Are they directly related to propagating longitudinal waves usually observed in ultraviolet in coronal magnetic structures like EUV fans \citep{2012RSPTA.370.3193D}? What is the nature of the recently discovered link between 3-min sunspot oscillations and the similar-periodicity flare oscillations \citep{2009A&A...505..791S}? An insight into these problems will give us a chance to diagnose plasma at different levels of the solar atmosphere, because waves carry the information on the nature of their emergence and on the properties of the medium where they propagate.

The review on dynamic of sunspots is arranged as follows: Section 2 addresses the spatial-frequency morphology of sources of sunspot oscillations and waves, including their localization,  size, oscillation periods, height localization with the mechanism of cut-off frequency that forms the observed emission variability. Section 3 presents a review dynamic of sunspot wave processes, provides the information about the structure of wave fronts and their time variations, investigates the oscillation frequency transformation depending on the wave energy. Section 4 addresses the initializing solar flares caused by trigger agents like magnetoacoustic waves, accelerated particle beams, and shocks. Special attention is paid to the relation between the flare reconnection periodic initialization and the dynamics of sunspot slow magnetoacoustic waves. Section 5 presents a short review of theoretical models of sunspot oscillations.

\section{Spatial structure of oscillation sources above sunspots}

Magnetoacoustic gravity (MAG) waves above sunspot atmosphere have a strong dispersion when propagating from the subphotospheric layers into the corona. Their properties dramatically depend on the cut-off frequency that, in turn, depends on the propagation medium plasma parameters. The waves with the frequency under the cut-off have a strong damping, and can not reach the upper layers of the solar atmosphere. Due to that, there occurs a filtration of the propagating broadband pulses by frequencies and height 
\citep{1982SoPh...75...99S,2011ApJ...728...84B, 2013A&A...554A.144Y}. The cut-off frequency in the stratified isothermal atmosphere with magnetic field was obtained analytically in \citep{1977A&A....55..239B, 1984A&A...133..333Z}. The frequency value in general case depends on the local plasma $\beta$ and on the magnetic field inclination angle $\varphi$. In the regions with a high plasma $\beta$ (photosphere and chromosphere of the quiet Sun), this frequency comes to the ordinary acoustic cut-off frequency, $\sim$ 5.2 mHz. For low plasma $\beta$
(sunspots, corona), the cut-off frequency is modified by the value of the inclination angles of power magnetic lines along which waves propagate. This explain the existence of low frequency compression waves in the corona 
\citep{2012RSPTA.370.3193D} that are assumed \citep{1993ApJ...405..787F, 2006ApJ...648L.151J} to transport the bulk of the energy as compared with high-frequency waves. The experimentally-obtained cut-off frequency can be used as a seismological instrument to study the local magnetic topology in sunspots.

In review \citep{2011LRSP....8....4B} the authors give an overview of the current state-of-knowledge of the magnetic
field structure in sunspots from an observational point of view. Briefly described  of tools that are most commonly 
employed to infer the magnetic field in the sunspot atmosphere. Taken into account the differences of global and local magnetic details of sunspots, focusing on the implications of the current observations of energy transport mechanisms for the different sunspots models. Presented different methods of extrapolations of the magnetic field towards the corona.

Investigations in \citep{1977A&A....55..239B} showed that long-period waves are capable of permeating through the photosphere and the chromosphere and of propagating upward, into the coronal region. The magnetic field concentration in the form of pores or sunspots with a complex magnetic structure are natural waveguides for these waves. The modification in the magnetic field line inclination is shown to be possibly an explanation for the origin of 5-min waves in chromospheric spicules \citep{2004Natur.430..536D} and coronal loops in active regions \citep{2012RSPTA.370.3193D, 2005ApJ...624L..61D}. One also observes the permeation of the solar global p-mode into the chromosphere through the inclined magnetic fields at peripheral portions of plages \citep{2009ApJ...702L.168D}. Subsequent observations of long-period oscillations in the corona were also explained by the existence of inclined magnetic channels along which waves propagate \citep{2009ApJ...697.1674M, 2011SoPh..272..101Y}.

	Direct observations of the magnetoacoustic cut-off in sunspots by using the spectropolarimetry results were first shown in \citep{2007ApJ...671.1005B}. The authors revealed that the cut-off frequency depends on the change in the magnetic field in approaching/approximation of a weak plasma beta \citep{1977A&A....55..239B}. However, there was no further investigation to determine how this frequency is related to the inclination of field lines. It became clear that the observed differences in the phase and correlation in the time series of the intensity variations obtained at different heights in the solar atmosphere look more realistic, taking into account the spatial distribution of the magnetic field and its curvature \citep{2007ApJ...671.1005B}.
	
\citep{2008ESPM...12.2.38K, 2011A&A...525A..41K} revealed a peculiarity in sunspot oscillations as a height inversion of power variation. At the chromosphere level, the 3-min oscillation power maximum spatial localization is usually over the umbra. At the photosphere level, these oscillations have a sharp damping. The revealed regularity agrees well with the results of the previous investigations. In particular, in \citep{1987SoPh..112...37B}, no 3-min periodicity components were observed. In \citep{1985ApJ...294..682L}, the level of the recorded 3-min oscillations was very low, at the noise level. By using helioseismology methods \citep{1990SoPh..129...83B,  2004SoPh..225..213N}, suppressing the level of photospheric oscillations in active regions was recorded. Analyzing the brightness fluctuations observed in the Hinode/SOT G-range 4305\r{AA} indicates at a decrease in the broadband brightness variations in the central part of sunspots \citep{2007PASJ...59S.631N}.

To explain the effect of changes in observed 3-min oscillations power with height two mechanisms was proposed \citep{2011A&A...525A..41K}. First of them connected with suppress the observed line Fe I 6569 \r{AA} at the photospheric level and decreases the amplitude of acoustic oscillations in deeper regions \citep{1992ASIC..375..261L, 1997A&A...324..743S} of sunspots. According to \citep{1983SoPh...88...71B}, the depression of the umbral photosphere with respect to the penumbra (the Wilson effect) is about 700 km. This difference in height can lead to the decrease in the 3-min oscillation amplitude by a factor of 2 \citep{1998ApJ...497..464L} and consistent with the 3-min the line-of-sight (LOS) velocity power distribution obtained at the photospheric level \citep{2011A&A...525A..41K}. The second mechanism connected with the topology of the sunspot magnetic field. In the umbra, the oscillations as a slow magnetoacoustic waves guided by the vertical magnetic field lines upwards and reach the chromosphere. At the umbra/penumbra border the magnetic field lines are inclined from the vertical direction and prevent the free propagation of the short period waves \citep{2006RSPTA.364..313B}. In \citep{2004ApJ...613L.185F} pointed out that the $\sim$ 2.4 min oscillations are reflected by the active region canopy. \citep{2011A&A...525A..41K} proposed that partial reflection of the waves from the magnetic dome can lead to the observed accumulation of the wave energy at the photospheric level outside umbra under horizontal magnetic field. This may explain the relative increase of the 3-min oscillation power outside the umbra in the photosphere, and the corresponding decrease of the chromospheric oscillations.

	In \citep{2006ApJ...647L..77M}, the authors studied the sunspot narrowband signal propagation time, and found good agreement with the study in \citep{1977A&A....55..239B}, both in the quiet Sun regions ($\beta$ $\gg$ 1) and in sunspots ($\beta$ $\ll$ 1). In \citep{2006A&A...456..689T}, the empirical formula $V_{\mathrm{peak}}(\varphi)\simeq 1.25V_{\max}(\varphi)$ \citep{2006RSPTA.364..313B} in the sunspot chromosphere was used to obtain the value of the magnetic field inclination. In this formula, $V_{\mathrm{peak}}$ denotes the peak frequency in the power spectrum, and $V_{\mathrm{max}}$ for the local cut-off frequency modified by the inclined magnetic field. In a recent study \citep{2012ApJ...746..119R}, spectral and phase variability in the emission intensity with height in ultraviolet was investigated. The existence of delays between the SDO/AIA temperature channels was shown, which indicates the presence of upward propagating brightness disturbances. The cut-off frequency height variations were found to agree well with the conclusions presented in 
\citep{1977A&A....55..239B}. Subsequent studies \citep{2012ApJ...756...35R} to compare the cut-off frequency of the observed magnetoacoustic waves at 304 \r{AA} (SDO/AIA) with the computed values of the extrapolated field showed good agreement with each other.
	
	In \citep{2013SoPh..284..379K}, direct measurements of phase delays between the wave speed variations above sunspots and faculae were performed. Different magnetic structures show different delay values related to the wave propagation speed in the selected region. The obtained speeds were revealed to dramatically surpass the generally accepted values of sonic speed in the photosphere. Thereupon, there arises the issue of these oscillations origin and of the wave mode recorded at observations.

\citep{2011SoPh..273..403A} compared the velocity oscillation modes at the photosphere-chromosphere level and the  radio emission oscillations (NoRH, 17 GHz) in the transition zone. The spectral composition of the wavelet spectra for optical and radio observations showed an about 50-sec time delay of signals. This indicates at the MHD wave propagation from below upward in a sunspot magnetic tube. 5-min oscillations show the same spectra at various heights only near the umbral region.

\citep{2013A&A...554A.146K} analyzed sunspot oscillations using the SDO/AIA and ground-based optical observation data. Low-frequency oscillations are shown to form sources at the sunspot penumbral boundary. Their shape agrees with the penumbral boundaries. At the coronal level, this component forms the magnetic fans' outer boundaries. 5-min oscillations are localized at the umbral/penumbral boundary. The fundamental 3-min component is inside the umbra. A higher-frequency component ($\simeq$ 2.1 min) is concentrated in individual small areas  of an umbra. The authors calculated the wave propagation velocities: 28 $\pm$ 15 km/s, 26 $\pm$ 15 km/s, and 55 $\pm$ 10 km/s for the Si I 10827\r{AA} - He I 10830\r{AA}, 1700\r{AA} - He II 304\r{AA}, and He II 304\r{AA} - Fe IX 171\r{AA} line pairs. Measuring the high rate of error indicates a lack of co-phase oscillations at different height levels due to their frequency dynamics.

	Observations of the photospheric magnetic field oscillations \citep{1999ASPC..184..136N, 1998SoPh..182...65B, 2000ApJ...534..989B} showed a fine field fragmentation in the regions with the maximal power of the oscillations coinciding with sunspots. For example, optical observations show that the magnetic field 5-min oscillations \citep{2000A&A...355..347Z} are localized both in the magnetic field isolated tubes (pores), and at the sunspot umbral/penumbral boundary. In sunspots, one also observes short-lived blinker-type flare events with a similar periodicity, 400 through 1600 sec \citep{1999A&A...351.1115H, 2000A&A...353.1083B, 2006A&A...450.1181L}.
	
	\citep{2014ARep...58..272K} studied efficiency of the oscillation mode permeation from the subphotospheric levels into the corona in the facular regions near sunspots. They used both the ground-based data in the Si I 10827\r{AA} and He I 10830\r{AA} lines and the SDO/AIA data  (Fe I 1700\r{AA}, He II 304\r{AA}, and Fe IX 171\r{AA}). The analysis of the spatial distribution in oscillation power at different heights showed that, at the coronal level (171\r{AA}), the emission oscillation maximum within the 11-17 min periods range is observed. This implies that the low frequency oscillations dominate in coronal loops over plages. 5-min oscillations that prevail everywhere in measurements of  velocities in the facula atmosphere lower layers are notable only in the detected compact parts of coronal loops. Further, \citep{2006SoPh..238..231K} revealed the spatial distribution of the sunspot velocity variations in a chevron form. These structures reflect the existence of propagating waves with a phase velocity of about 50-80 km/s. The comparison between the characteristics of the 3-min umbral waves and the penumbral traveling waves at the chromosphere level revealed no association between them. A possibility to explain the oscillation frequency and the propagation velocity depending on the umbral center within the interaction and transformation of different oscillation modes is discussed. For the traveling 5-min penumbral wave, the velocity is shown to insignificantly vary with the distance from the umbral border.
	
	The spatial distribution of the sunspot oscillation frequency and power has a different degree of insight \citep{2006RSPTA.364..313B}, and was vigorously studied by using the data obtained with both ground-based, and space instruments (SOHO, TRACE, HINODE). These studies, the SOHO/MDI observations in particular, showed that, at the photosphere level, there occurs an increase in the Doppler velocity and a line broadening, but not the continuum intensity \citep{2002A&A...387.1092J}. An increase in the spectral power was discovered in active regions with a sufficiently strong magnetic field. In turn, as the field further strengthens, a significant decrease in oscillations is observed, for example, in the sunspot center. The observations in H$\alpha$ showed that steady oscillations (with the period of about 2.6 min) dominate in the umbral region and in the inner part of sunspot penumbra, with the presence of traveling waves in the penumbra within the 5.6 min range. However, these waves were not recorded at the photospheric heights \citep{2000A&A...354..305C, 2001A&A...375..617C}. The sunspot umbra 3-min oscillations and the penumbral 5-min traveling waves were assumed to show an example of reflecting the dynamics of the same waves generated in the photosphere and propagating along different magnetic field lines with a different inclination level \citep{2007ApJ...671.1005B}. On the other hand, in \citep{2009Ge&Ae..49..935K}, no association was revealed between the umbral 3-min oscillations and the penumbral 5-min traveling waves at the chromosphere level. Parameters for the 5-min oscillation mode were obtained. The length of propagating waves is $\sim$12-30 arcsec, or 9-20 Mm, the phase velocity varies from 28 to 65 km/s.
		
	\citep{2003ApJ...591..416C} showed that oscillations with different frequencies within the 3-min period range are localized in different parts of an umbra at the chromosphere level. One observes numerous oscillation peaks in the 2-3 min spectral range. There are three detected harmonics in the spectrum: the largest is near 3 min, the second is near 2.6 min, and the third (about 2.2 min) is present for a restricted time. Within existing theories, the $\sim$3 min oscillations are considered as an effect of the photospheric resonator, two other modes are considered within the chromospheric resonator.
	
	In solar atmosphere take place the formation of acoustic wave resonators both in active and quiet Sun regions \citep{1994ApJ...426..404S, 2002ApJ...564..508R, 2008A&A...481L..95S}. For example the analysis of observational data, obtained by Hinode/EIS \citep{2008A&A...481L..95S}, show $\sim$ 7-13 min intensity oscillations (He II 256.32 \r{AA}, FeXI 188.23 \r{AA}, and FeXII 195.12 \r{AA}) in brightest core of the magnetic network. The authors suggest that the field-free cavity regions, under the bipolar small-scale magnetic canopy may serve as resonators for the waves. The field-free cavities probably have granular dimensions. The MHD equations in the cavity and canopy regions was solved separately, and merge the solutions at the cavity-canopy interface. The dispersion relation for oscillations shows that the first harmonic is leaky, and that the oscillations may propagate upwards. Numerical modeling of the wave propagation through magnetic structures in a stratified atmosphere \citep{2002ApJ...564..508R} shown, that in regions with significantly inclined field to the solar normal, refraction by the rapidly increasing phase speed of the fast modes results in total internal reflection of the waves. The height of this reflecting layer will depend on the field geometry and wave frequency, but in the Sun it will certainly be a highly complicated surface. This varying reflection height may be responsible for the intermittancy seen in the oscillations. \citep{2009A&A...505..763K} studied the influence of this small-scale magnetic canopy, on the wave dynamics. They found that the field-free cavity regions, under the canopy, may trap high-frequency acoustic oscillations. The lower-frequency oscillations may be channeled upwards in the form of magneto-acoustic waves.  

\citep{2006A&A...456..689T} found that the 3-min oscillation frequency in the chromosphere dramatically changes at the umbral/penumbral boundary when moving from the sunspot center. Revealed were small bright mottles in the sunspot center where the oscillations with a period less 4.1 min tended to decrease \citep{2007PASJ...59S.631N}. A similar effect was found on the Doppler velocity maps in the Ca II 8542\r{AA}  line \citep{2007A&A...463.1153T}. The nature of these "dark spots" in the spatial distribution of 3-min oscillations at the chromosphere level still is not clear.

\begin{figure*}[t]
\label{fig1}
\epsfysize=8.5cm \centerline{\epsfbox{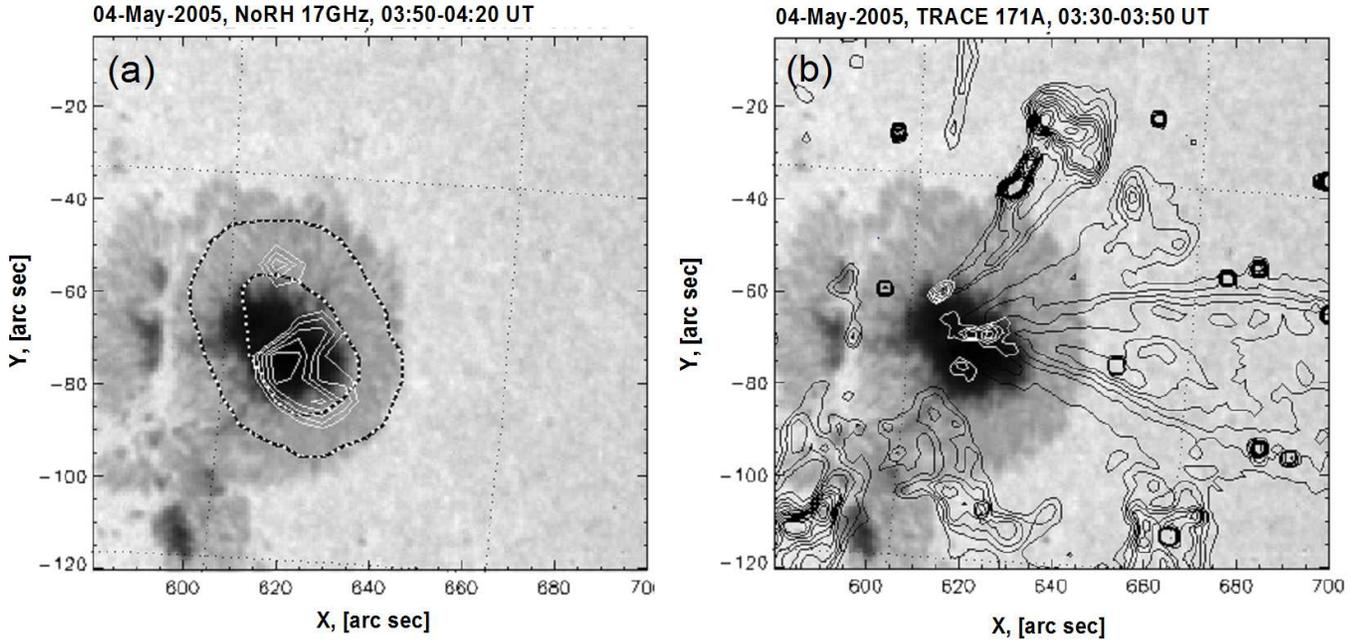}}
\caption{{\it Left panel:} V-shaped wakes (white contours) from propagating 3-min waves in the radio-frequency range (NoRH, 17 GHz). {\it Right panel:} EUV narrowband 3-min wakes in the corona (white contours). Black contours designate the radio source and coronal loops in the broadband emission. The background is the sunspot image in the white light.{\bf \citep{2010SoPh..266..349S}}}
\end{figure*}

\citep{2013ApJ...779..168J} studied the role of the magnetic field inclination angles in forming physical characteristics of traveling waves in a penumbra at the chromosphere level. An increase in the period oscillation depending from the umbra center was found. Based on applying the cut-off frequency mechanism, a possibility to obtain the magnetic field structure above sunspot by using the data on  dominating wave periods' spatial distribution was shown. It is concluded that there is a relationship between the geometry of the magnetic field connecting the photosphere with the chromosphere and the behaviour of traveling waves in a penumbra in the upper chromosphere. This conclusion directly supports the statement that the phenomenon of traveling waves in a penumbra is a chromospheric response to the upward propagating magnetoacoustic waves generated in the photosphere.

	For transition zone using a  short-baseline radiointerferometer (SBRI) \citep{1983IGAFS..65...97Z, 1984SoPh...93..301Z} showed the existence of oscillation sources, distributed in space with 3-, 5- and 7-minute periods. Further, using one-dimensional observations at the Siberian Solar Radio Telescope (SSRT, 5.7 GHz) \citep{1989IGAFS..87...113Z, 1992IGAFS..98...114Z}, as well as the Nobeyama radioheliograph (NoRH, 17 GHz) two-dimensional images \citep{1999SoPh..185..177G}, revealed were the radio brightness variations leading not only to a brightening of the part of the source coinciding with a sunspot umbra, but also to periodic spatial displacements. This indicates at the existence of a fine spatial structures inside an active region in the form of oscillating small-angular-size sources that are not spatially resolved by a radiotelescope.
	
	Further investigations \citep{2001ApJ...550.1113S} by using correlation curves and two-dimensional images of radio sources showed that the 3-min oscillations steadily exist in sunspots over long periods. Applying the density and temperature values obtained at simultaneous observations with SOHO/SUMMER to calculate the gyroresonance emission, found was a good agreement with the parameters of the detected radio emission. It was concluded that brightness variations in sources may be caused by density and temperature oscillations when acoustic waves pass upward through the third gyroresonance level with the corresponding mean value of the magnetic field ($\sim$2000 G). In this study, integral radio images of sources were used, which did not allow one to obtain their (sources) fine spatial and frequency structure. 	\citep{2004AstL...30..489G} suggested that the periodicity might be explained by the existence of resonance structures for acoustic and 
Alfv\'en waves. The position of the resonators, as well as their size, may change within rather broad limits and, correspondingly, generate different observed oscillation periods. Three basic resonator types have been proposed: the resonator coinciding with the emission region, the resonator outside the emission region, but quite close to the radio source, and the resonator of the global solar nature, similar to 5-min oscillations.

\citep{2002A&A...386..658N} used investigation into difference radio maps obtained with the VLA radiotelescope in the two wavelengths, 8.5 and 5 GHz, with a high spatial resolution. The emission oscillations from a sunspot are shown to indicate sharp variations in brightness with a fast rise and slow drop. Despite such variations, they feature spatial, amplitude, and phase stability. The intensity variation spatial distribution is presented like fine-structure details of a small angular size at the umbral boundary. These structures have a circular distribution as a patches, and their emission arrives from the gyrofrequency second level.

\citep{2008SoPh..248..395S} studied spatial, temporal, and phase dependences of oscillations in the sunspot-related radio sources (NoRH, 17 GHz). Because the instrument resolution is not sufficient for direct spatial resolution of wave, the Pixelized Wavelet Filtration method  (PWF-analysis) was developed. This numerical method is based on the signal decomposition for each pixel of the image cube into individual spectral components by using wavelet transform. This allowed one to obtain not only oscillation power spatial distribution for individual harmonics in the oscillation spectrum (narrowband sources), but also to trace their time dynamics. The 3-min sunspot oscillations are localized in the center of microwave sources related to sunspot umbra. The 5-min component is localized in the small-size symmetric details at the umbral/penumbral boundary. This location is similar to that observed with VLA \citep{2002A&A...386..658N}, taking into account the difference in the spatial resolution of instruments. There are signal phase differences depending on the spatial localization of details.

\begin{figure*}[t]
\label{fig2}
\epsfysize=13cm \centerline{\epsfbox{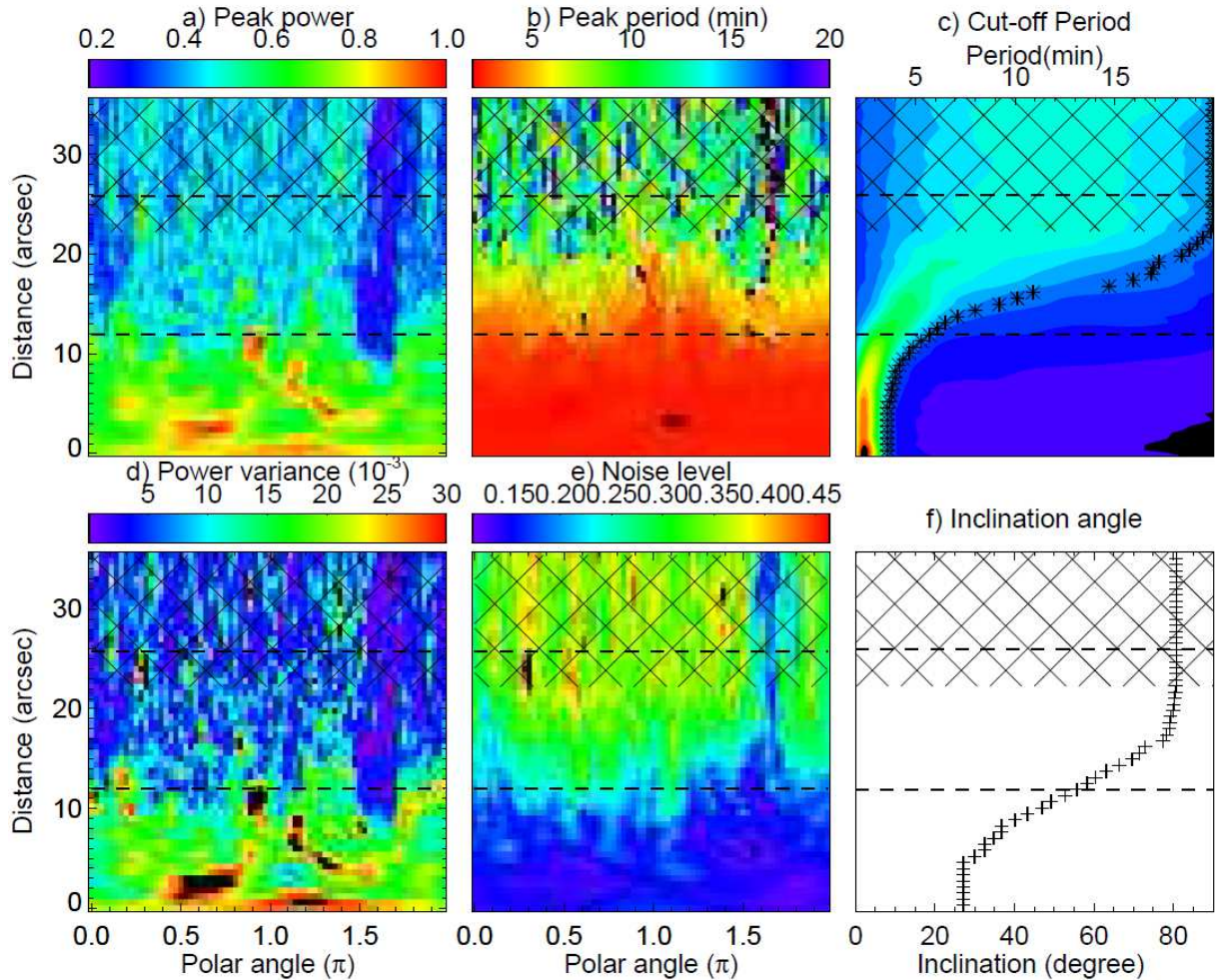}}
\caption{The oscillating power information extracted from the 304 \r{AA} data set of the sunspot AR11131 (08 Dec 2010), shown in the polar coordinates. The dashed lines mark the borders of the umbra and penumbra. The unreliable region is cross-hatched. a) The peak power distribution. b) The peak period distribution. c) The 1D spectral power map as a function of distance and period. The contour in asterisk shows the cutoff period. d) The spectral power variance distribution. e) The spectral noise distribution. f) The reconstruction of the magnetic field inclination.{\bf \citep{2014A&A...561A..19Y}}}
\end{figure*}

	\citep{2010SoPh..266..349S} found that the 3-min oscillation power distribution in the radio sources (NoRH, 17 GHz) like a patches inside umbra and coincide with the foot points of open magnetic loops in the TRACE 171 \r{AA} ultraviolet range (Fig.1). The
emergence of these spatial details in the morphology of the 3-min wave traces can be interpreted as the non-stationarity of the wave leakage from the sub-photospheric regions to the transition region. The wave traces are found to have round and extended shapes. This indicates the change of the wave-vector component in the plane perpendicular to the LOS. In the quiet stage of sunspot activity, the wave traces had a round shape, indicating that the waves were propagating mainly along the LOS or were standing in the LOS direction. As the LOS was almost perpendicular to the solar surface in the vicinity of the sunspot, this corresponded to the vertical motions of plasma. The periodical increases $\sim$ 10-30 min in wave disturbances can result the emergence of V-shaped wakes and correspond to the horizontal  wave propagation in foot points of inclined magnetic tubes in the corona. The fine structure of the three-minute wave traces at the transition region and coronal levels is found to almost coincide. Thus, it is likely that these phenomena are caused by the same wave motion that propagates from the deeper sunspot regions into the corona. For 3-min oscillations, there exists a correlation between the increase in the wave amplitude and the increase in the horizontal projected phase speed. The change of the speed is likely to be attributed to the change in the magnetic-field geometry, perhaps caused by the waves. In the transition region the growth of the 3-min amplitude is accompanied by the decrease in the LOS angle to the wave-propagation direction. In other words, the higher amplitude waves are observed to propagate more vertically. A possible interpretation of this phenomenon is the effect of the centrifugal force produced by the field-aligned flows of plasma along the curved magnetic-field lines \citep{2000A&A...362.1151N, 2006RSPTA.364..461D, 2009SSRv..149...65D}.
			
	For high-frequency oscillations, the sources with the oscillation period less than 3 minutes are localized in the umbra and decrease in size as the period decreases \citep{2008SoPh..248..395S, 2014A&A...561A..19Y, 2012ApJ...757..160J}. As the period grows, the size of sources also increase. At the chromosphere level, they reach the maximal value with the oscillation period of about 20 minutes, localizing at the penumbral boundary. For the symmetric sunspots, the sources of low-frequency oscillations with the period more than $\sim$ 3 minutes look like expanding rings, resembling the umbral boundary in shape. Small bright patches of the 5-min oscillation sources in the radio-frequency range (NoRH, 17 GHz) \citep{2008SoPh..248..395S} and (VLA, 8.5 GHz) \citep{2002A&A...386..658N} may merge, forming circles \citep{2014A&A...561A..19Y, 2013SoPh..284..379K, 2000A&A...355..347Z}. The localization of the sources at the umbral/penumbral boundary indicates at a strong coupling between acoustic waves and the magnetic field in the region, where local conditions favor the absorption of the global solar p-mode \citep{2003MNRAS.346..381C, 2006MNRAS.372..551S}. The observed spatial dependencies of oscillation sources can be explained by the spatial distribution of the magnetic field and its curvature in the framework of modification of the cut-off frequency 
\citep{1977A&A....55..239B}.

\citep{2012ApJ...746..119R} found that the SDO/AIA ultraviolet high-frequency oscillations within the 1.8-3.3 min range are mostly pronounced closer to the umbral center, while the lower-frequency component in the form of oscillating rings concentrates on peripheral sites of the umbral/penumbral boundary. Similar spectral characteristics of oscillations were found for all the temperature channels. The oscillation region slightly expands with height. The oscillating ring diameter broadens by about 1 arcsec with each subsequent SDO/AIA temperature channel, overstepping the boundary at 304\r{AA}. In the 171\r{AA}-211\r{AA} coronal channels, the high-power oscillations are localized in coronal fan structures. Higher frequencies concentrate toward the center at 1.8-2.1 min range, but their power decreases. The cut-off frequency is correspondingly observed to decrease from 5.7 mHz to 5.0 mHz or increase from 2.9 min to 3.3 min in period range. This discovery is interpreted as a manifestation for the variation in the inclination of the magnetic field lines that exit the sunspot at the temperature minimum level. This confirms the     assumptions that sunspot strong magnetic fields play the role of waveguides for acoustic waves propagating upward from the subphotospheric levels into the corona.

\citep{2012ApJ...756...35R} calculated the magnetic field in potential approximation and studied the distribution of the oscillation power in the sunspot atmosphere from the SDO/AIA data. The authors shown that the propagating waves are manifestations of a common oscillatory phenomenon, like the leakage of the solar global p-mode from the photosphere \citep{2005ApJ...624L..61D}. The size of the 3-min oscillation sources localized in the umbral region was found to grow with height as magnetic field lines expand. There is transformation of the oscillation period from 3-min in the umbra to 5-min traveling waves in the penumbra. This effect may be explained by the frequency variation in the sunspot region. To interpret this variation, the authors have to refer to the theory of the magnetoacoustic gravity (MAG) waves, that is, the acoustic-like slow-mode propagation along magnetic field lines. Due to the presence of a gravitational field, compressible MAG waves can propagate upward only at frequencies above the cut-off frequency with effective gravity dependence. In regions of low plasma $\beta$, which is the case above the sunspot umbra, the effective gravity on a particular magnetic field line should be decreased by the cosine of the inclination angle $\Theta$ of that field line with respect to the direction of gravity. Therefore, variation of the cut-off frequency across the sunspot and observed spatial-frequency shape of oscillation sources accordingly, depends only on the change of the field inclination angle.

The study \citep{2014A&A...561A..19Y} has become a further investigation into the cut-off frequency effect on the distribution of the oscillation source shape in the polar coordinates (Fig. 2). The oscillations up to 20 minutes were studied. The short-period oscillations were shown to concentrate within the umbral boundaries. The long-period sources of oscillations are located in the penumbral region as expanding rings. There exists a dependence between the oscillation power peak in the Fourier spectrum and the source boundary distance to the sunspot center. The lowest-frequency oscillations concentrate at the penumbral boundary. The magnetic field was reconstructed based on the MAG wave theory and spatial distribution of the cut-off frequency. A good agreement was found with the parameters of the extrapolated magnetic field in potential approximation. The obtained values for the inclinations of the reconstructed fields are mainly higher, than those in the case of the extrapolated field. This discrepancy assumes a significant effect of other physical processes affecting propagation of waves, such as radiation cooling that can the change in cut-off frequency \citep{2006ASPC..358..465C, 2009ApJ...692.1211C, 2010ApJ...722..131F, 1983SoPh...87...77R, 1991A&A...241..625U, 2010MNRAS.405.2317S} and MHD mode conversion \citep{2009ApJ...694..573P, 2011ApJ...738..119C, 2012ApJ...746...68K}. The developed method is alternative for the field reconstruction, and may be applied to specify the boundary and/or initial conditions when using the routine methods for the magnetic field extrapolation.

Proceeding from the cut-off frequency distribution over a sunspot, one may assume that narrowband sources with a period less than 3 minutes will also have a ring-shaped structure like lower-frequency components \citep{2012ApJ...746..119R}. Inside the umbra with the magnetic field vertical component, a slow variation in the field spatial inclination relative to the solar normal line occurs. The cut-off frequency, as a cosine of the field line inclination angle, slowly changes. Correspondingly to these changes, slow variations in the diameter of ring-shaped sources will occur in the 3-min oscillation band with a continuous filling of the umbra. For the penumbra with a fast change in the field inclination, the rate of the change in the source diameter increases, and the detected rings, without superposing on each other, are correspondingly observed.

\begin{figure*}[t]
\label{fig3}
\epsfysize=8.2cm \centerline{\epsfbox{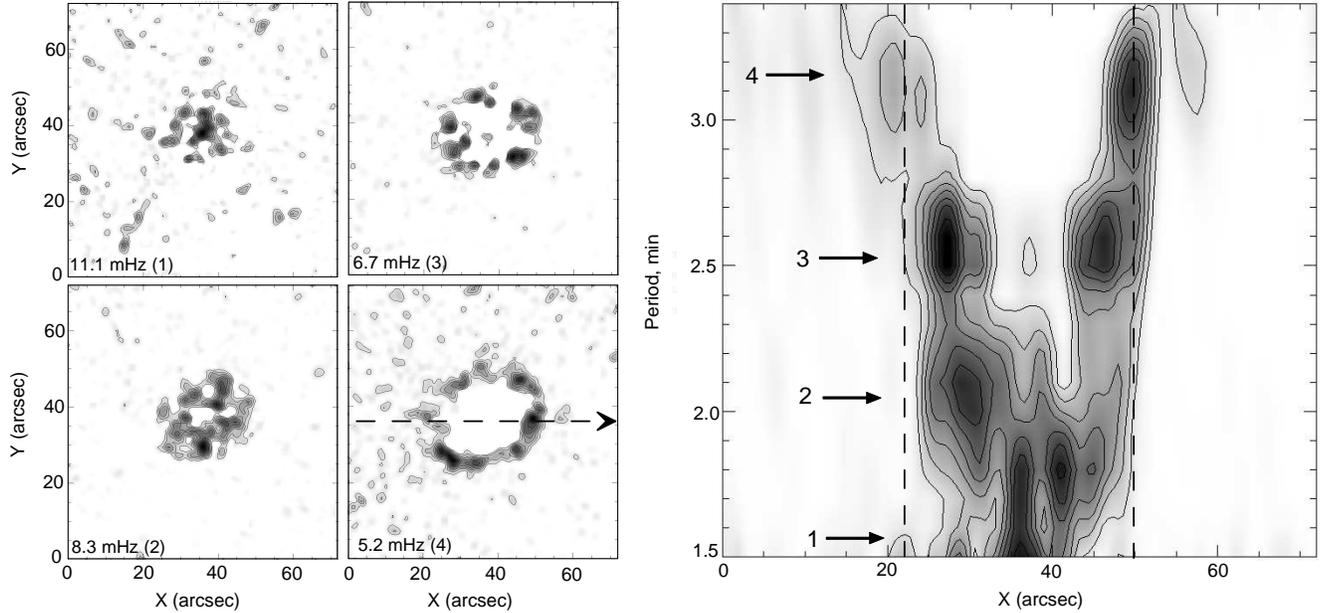}}
\caption{{\it Left panel:} Running-difference narrowband maps of the umbral oscillations signals obtained at 304 \r{AA}, integrated from 02:30 to 03:30 UT, at the frequencies 11.11 mHz (1.5 min), 8.3 mHz (2.0 min), 6.7 mHz (2.5 min), and 5.2 mHz (3.2 min).  {\it Right panel:} Dependence of the narrowband power distribution across the umbra (in the horizontal direction via the sunspot center, see the dashed arrow in the {\bf $\sim$3 min} map in {\it Left panel}) on the oscillation period. The vertical dashed lines show the umbral boundaries. The digits indicate the peaks in the Fourier spectra.{\bf \citep{2014A&A...569A..72S}}}
\end{figure*}

	\citep{2014A&A...569A..72S} obtained in first time the evidence of a fine spatial structure of oscillation sources in the 1.5 - 3.5 min oscillation band, and  compared them to the Fourier spectrum harmonics. The authors performed spatial spectral decomposition of the umbra in the 3-min oscillation band by using the PWF-analysis \citep{2008SoPh..248..395S}. The observation duration was one hour. Figure 3 presents the obtained results in the form of narrowband difference maps for the detected narrowband harmonics (Fig. 3, {\it Left panel} ), and their spatial spectral dynamics on the time-distance plots (Fig. 3, {\it Right panel} ). Scanning the narrowband source brightness variation occurred through the umbral center. Fig. 3, {\it Left panel} shows that the high-frequency oscillations with the $\sim$1.5-min period (the beginning of the 3-min band) mainly concentrate in the center. As the periods grow, the rings form, whose diameter variations create V-shaped diverging branches (Fig. 3, {\it Right panel} ). A minimal filling (point high-frequency source) will be observed at the beginning of the 3-min band. The umbra filling maximal value will occur at the end of the band. This results in the effect of the frequency rings' nesting in one another, and in forming continuous filling of the umbra, which is observed for the 3-min band oscillation sources \citep{2012ApJ...746..119R}. The oscillation power variation irregularities along the branches agree with a fine-structure distribution of spectral harmonics in the 3-min oscillation band of the integral Fourier spectrum. In the penumbra, the inclination of magnetic field lines sharply increases, which leads to acceleration of the ring diameter increase for the lower-frequency components.
	
	\citep{2014A&A...561A..19Y} analyzed 5-min oscillations in the region of umbral light bridges by using the HINODE/SOT instrument. This component is shown not to vary dramatically along the bridge, which indicates at the oscillation emergence beneath. The obtained oscillation characteristics are similar to those of the penumbral 5-min traveling waves. The 3-min oscillations are shown to be localized in all the umbral mottles separated by light bridges. The opposite sides of bridges oscillate in phase, which indicates at a uniform source of 3-min oscillations for different mottles. Umbral flashes are discovered to be a continuation of the umbral oscillations, with a larger amplitude, and with no effect on the fundamental oscillation phase. The oscillation period is about 3 minutes.

\subsection{Summary}

      There is a dependence revealed between the spatial-height distribution of the cut-off frequency in the sunspot atmosphere and the fine structure of the oscillation sources in narrowband images. The oscillation sources with periods less than 3 minutes are shown to be located in an umbra, decreasing in size as the cut-off frequency increases. The 5-min oscillation sources are localized around the umbra-penumbra boundary. With the period growth, the size of the sources increase, reaching the penumbra boundaries. This regularity reflects the dynamics of the broadband pulses generated at the subphotospheric level and propagating in diverging magnetic field lines with different inclination to the solar normal and, correspondingly, with a different cut-off frequency. In the sunspot atmosphere high layers, there form the wakes seen like V-shaped brightenings at the coronal loop foot points. The detected dependences show that sunspot strong magnetic fields play a role of waveguides for the acoustic waves propagating into the corona. The observational values of the frequencies for different layers of the solar atmosphere allow obtaining both one-dimensional (in the source plane) and two-dimensional (by height) distribution of the field line inclination angles and, correspondingly, reconstructing the magnetic field by using the helioseismology method.

\section{Wave dynamics in sunspot}

The chromosphere-corona transition region is a very variable zone that is not steady in its height position and size. Substance motions are observed both up and down. Along with variations in the fine spatial localization of sources, \citep{2003A&A...410..315R} revealed sign-alternating changes in the form of the observed wave frequency drifts in sunspot details (plumes) at the SOHO/SUMMER spectrograph in the CIV, NeVIII lines, and in the EIT ultraviolet radiation (171\r{AA}). Most oscillations in the transition region and lower corona concentrate near 2.7 min, with its amplitude modulation in the wave train form, 10-20 minutes (sometimes up to 40 minutes) long.  Variations in the Doppler velocity by the line-of-sight and intensity arise in the same period interval (2.3-5.5 min) with a pronounced drift toward short periods. The observed frequency drifts of velocity oscillations in the chromosphere as well as in the chromosphere-corona transition region may be explained in framework of model assuming a parallel stratification of sunspot atmosphere with a shrinking resonator size. The observability and the line-of-sight component of the Doppler velocity depend on the direction of the magnetic field. In expanding sunspot magnetic field the oscillating volume may change its direction relative to the line of sight while moving along the field lines and thus cause a variation of the observed velocity. Oscillations also can reach regions of considerably converging flux tubes and provides additional changes of frequency.

\citep{2006ASPC..358..465C,2009ApJ...692.1211C}, by using linear wave equations with radiation cooling, showed good agreement between the observed phase delays and wave amplitude height variations, both for sunspots and for pores. A detailed study for a larger number of the emission lines emerging at different heights is presented in \citep{2010ApJ...722..131F}. Phase differences of propagating waves and an increase in their power are shown to be interconnected. Therefore, studying the cut-off frequency may also provide information on plasma parameters related to radiative losses.

\citep{2001A&A...368..639F}, using the SOHO/CDS data, showed the existence of significant frequency drifts in the 2.8 min oscillations above sunspots in the lines of the transition region. The typical train duration is about 15 minutes. Different spaced sources correspond to different frequencies in the spectrum. There are rare instants when the details with oscillations at several frequencies exist simultaneously.

\begin{figure*}[t]
\label{fig4}
\epsfysize=9.5cm \centerline{\epsfbox{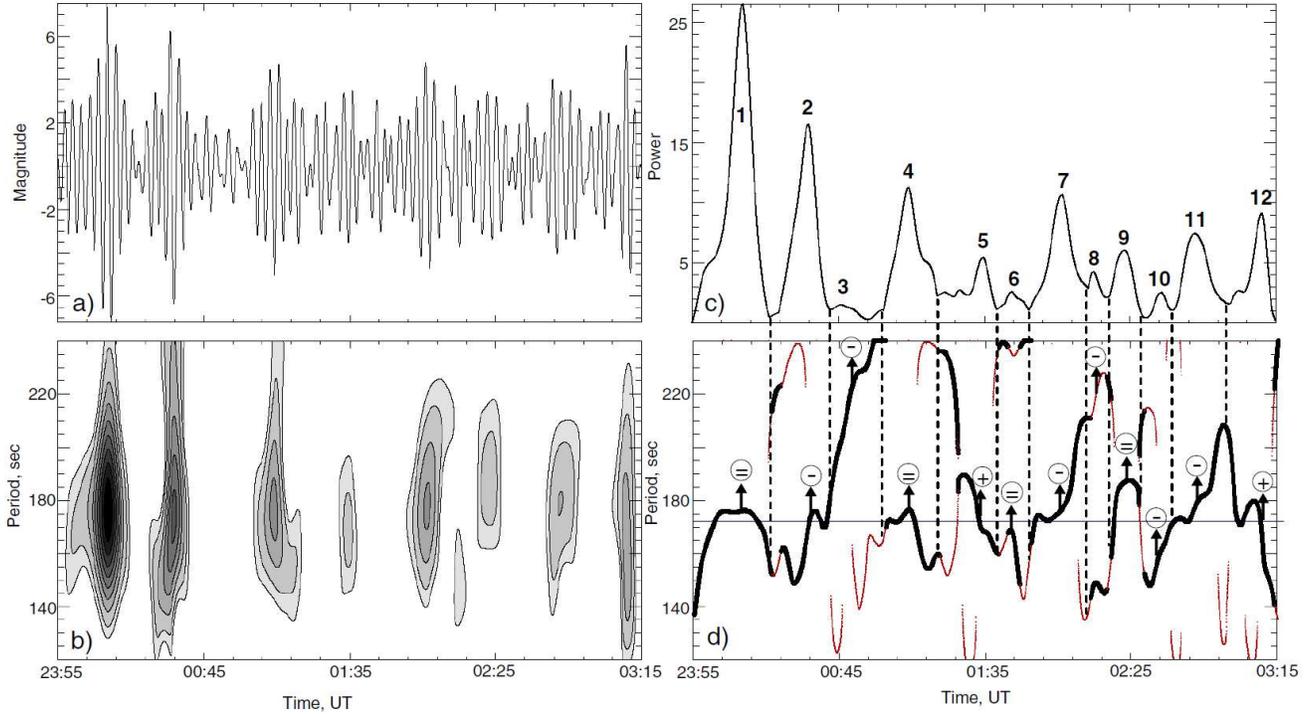}}
\caption{Wavelet analysis of the 3-min oscillation of the microwave signal on May 3, 2005, from 23:55 to 03:15 UT. Panel a) shows the amplitudes obtained by calculating the inverse wavelet transform of the signal in the 120-240 s period range. Panel b) power distribution of the 3-min oscillation trains in the wavelet spectrum. Panel c): time evolution of the 3-min oscillation power. The numbers show the numeration of the trains. Panel d) shows the wavelet skeleton representing both global (thick lines) and local (thin lines) extrema. In the circles we show the signs of the period drifts (positive, negative or without drift). Time is shown in UT, power and amplitudes are in arbitrary units.{\bf \citep{2012A&A...539A..23S}}}
\end{figure*}

\citep{2003ApJ...591..416C}, by using wavelet spectra, showed the existence of the 3-min oscillation dynamics. There are frequency drifts observed, oscillation modes tend to change, to attenuate or augment over short time intervals. This behavior was explained by both the connection between closely frequency-spaced oscillation modes and variations in the physical properties of resonators.

	\citep{2012A&A...539A..23S} showed that the sunspot-related 3-min oscillations in radio sources (NoRH, 17 GHz) have a non-monotonic character. There exist low-frequency oscillation trains with an $\sim$8-20 min period ($\sim$13 minutes being the mean period value). The interval between trains is $\sim$13-60 minutes. The oscillation trains vary both in power and in frequency. The signal variation relative amplitude is $\sim$3-8$\%$. Pronounced signal frequency drifts are observed throughout the evolution of individual oscillation trains. The beginning and the end of the drifts agree with the beginning and the end of the amplitude modulation. The period changes from 90 to 240 seconds. The wavelet analysis showed the existence of both positive and negative frequency drifts. The frequency drift velocity for the 3-min fundamental harmonic throughout the low-frequency modulation tends to increase with height, reaching 4-5 mHz/hr in the photosphere, 5-8 mHz/hr in the chromosphere, and 11-13 mHz/hr in the corona. Figure 4 presents an example of the 3-min oscillation frequency drifts in the radio-frequency range.
	
There exist significant spatial variations in narrowband 3-min sources throughout the frequency drift. New oscillation trains and frequency drifts in the transition region are related to the emergence of new oscillation sources in the form of V-shaped structures. These changes are noted at all the heights of the sunspot atmosphere. When spatially comparing the oscillation sources with observations in the 171\r{AA} line (SDO/AIA), obtained is their coincidence with the foot points of coronal arches. These structures may be interpreted as the emergence of new waveguides along which waves propagate upward. A possible explanation for the dynamic properties of the observed waves is in the effect of the two factors: dispersion evolution of the wave pulses propagating upward, and inhomogeneity of the oscillation power distribution over an umbra. The evolution of initial broadband pulses results in forming oscillation traces \citep{1982ApJ...256..761R}. Because the sunspot cut-off period is about 3 minutes, the observed extended oscillation sources may be explained as the traces behind the pulses propagating along the magnetic field \citep{1991A&A...250..235F, 1994A&A...284..976K, 1995A&A...294..232S, 2009A&A...505..763K}. Different magnetic waveguides have different physical (temperature, density) and geometrical (inclination, length, width) parameters. Correspondingly to the change in the oscillation power and the train emergence, waves propagate through different magnetic waveguides forming a unique cut-off frequency. The maximum power prevalence in a detected tube determines the instantaneous frequency. The power variation and, correspondingly, the oscillation frequency are related to the propagation along different magnetic tubes. A sequence of similar variations forms frequency drifts.

	There exists an ambiguous opinion concerning the interconnection between the wave processes in plumes and coronal fan-structures with the 3-min oscillations. \citep{2004SoPh..221..237B} showed that the umbral oscillations are confined to small regions related to foot points of coronal loops. There was an assumption about the absence of relation between oscillations and plumes. However, \citep{2012ApJ...757..160J} found direct agreement of 3-min oscillations with coronal fans that are anchored in the umbra. Their foot points are photospheric umbral dots where the oscillation power amplification is observed. \citep{2009ASPC..415...28W} recorded 12- and 25-min oscillations of intensity and radial velocity in fan like coronal structures over active regions using the Hinode Extreme-Ultraviolet Imaging Spectrometer (EIS) data. These oscillations were identified as propagating slow magnetoacoustic waves.
	
	Till present, there is no sufficient information on the wavefront dynamics at different levels of the solar atmosphere and their stability over long periods. When observing at a moderate spatial resolution in the 17 GHz range \citep{1999SoPh..185..177G}, the 3-min oscillations were noted to be mainly localized at the radio source center, and look like periodically repeating brightness increases. Further, at increasing the temporal and spatial resolutions (TRACE, SDO/AIA), the obtained movie and time-distance plots clearly indicate at the difference between the wave fronts at the bottom and at the top of the solar atmosphere. At the lower layers of solar atmosphere, there is a quasi-spherical propagation of wave fronts with the center in sunspot umbrae \citep{2010SoPh..266..349S}. For the coronal level, along with the radial propagation along low magnetic loops, there emerge individual detected directions coinciding with the wave propagation along high, open field lines. Their investigations done by using one-dimensional data in the form of scanning the active region with a spectrograph slit \citep{2006SoPh..238..231K} allowed one to reveal one-dimensional symmetric chevrone-type wave fronts on time-distance plots. The front structure shows spatial variations with time, which indicates at the variation in the wave propagation velocity along a sunspot.
	
	\citep{2014A&A...569A..72S} analyzed the dynamics of one- and two-dimensional wave fronts. There exist instants on the one-dimensional time-distance plots, when a significant departure from parallelism between propagating wave fronts is observed. There is direct correlation between 1D wave form and the oscillating signal level. At the oscillation maximum, the waves have a symmetric, spherical shape on either side from the sunspot center. At a decrease in the signal level, a decrease in symmetry occurs, fractures and spatial displacements with front shape breakup are observed. A two-dimensional PWF-analysis \citep{2008SoPh..248..395S} of the wavefront trains in the 3-min period band showed that, for the maximal power oscillation trains, characteristic is the wave propagation in the form of a spiral with two branches fixed in a pulsing point source at the umbral center. A counterclockwise rotation of the spiral with a successive expansion of the branches up to the shadow boundary is observed (Fig. 5). As the power decreases, a partial collapse of the branches into individual components occurs. Quasi-spherical waves form. A similar two-dimensional dynamics is confirmed by good correlation with one-dimensional time variations on the time-distance plots. One should note that the 3-min peak has a certain spectral band. Therefore, taking its fine structure and the presence of subharmonics into account, one may expect differences both in the shape of narrowband oscillation sources and in their dynamics. The analysis of fine frequency structure of the wave source shape in the 3-min umbral oscillation band showed that the observed quasi-helicity is an assemblage of narrowband details with different frequency, spatially separated over the umbral space. In the umbral central part, a high-frequency pulsing point source with a period of $\sim$1.8 min connected to the lower-frequency arc-wise expanding details within the 2.6-3.1 min period range.
	
\begin{figure*}[t]
\label{fig5}
\epsfysize=9cm \centerline{\epsfbox{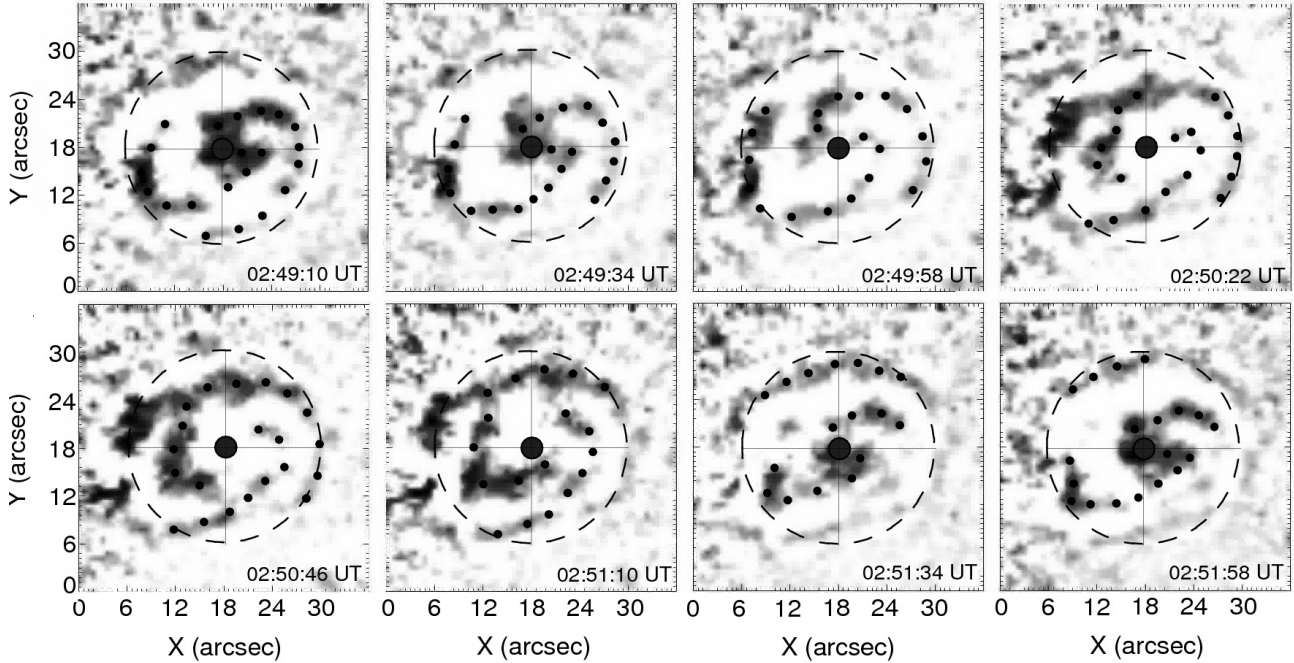}}
\caption{Temporal dynamics of the narrowband wave fronts throughout one cycle of 3-min umbral oscillations at the transition region level (SDO/AIA 304 \r{AA}). Black dots indicate front borders. The dash circle shows the umbral boundary.
{\bf \citep{2014A&A...569A..72S}}}
\end{figure*}
	
	The observed global horizontal oscillations may be interpreted within the model based on the frequency cut-off of propagating waves. According to this model \citep{2008SoPh..251..501Z, 2011ApJ...728...84B, 2012A&A...539A..23S}, the oscillation frequency decreases with decreasing of the magnetic waveguide inclination angles relative to the solar normal line. In the sunspot center where the field is vertical, oscillation sources should have a symmetric shape and a maximal value of the cut-off frequency \citep{2014A&A...561A..19Y} which is observed in reality. The magnetic field shows no initial helicity \citep{2012ApJ...756...35R}. One may assume that the magnetic field lines anchored in one place have allocated spatial sectors, where the inclination angles will be different in the polar coordinate system. In this case, the circular symmetry of wave fronts will be violated with the formation of individual segments in the form of the arc-wise frequency sources in the 3-min band. The sequential location of these frequency sources from the sunspot center forms the observed quasi-helicity.
	
\subsection{Summary}

   The solar sunspot emission dynamics in the broad frequency band shows a well-defined low-frequency modulation of 3-min oscillations in the form of trains. The modulation is non-stationary by both the power and the frequency. There are frequency drifts over trains. The start and the end of the drifts coincide with the start and the end of the amplitude modulation. The 3-min oscillation drift rates were found to grow with height. The drift start coincides with the emergence of fine-structure oscillating details with the maximal power at the foot points  of coronal arches. The observed spatial-time-frequency peculiarities of the oscillations may be treated within two mechanisms operating simultaneously. The first is related to the dispersion evolution of the subphotospheric pulses propagating upward along the sunspot magnetic field. The other mechanism is related to a spatial split of the propagating pulses into several individual magnetic tubes with different physical and geometric parameters. For low heights, characteristic is a quasi-spherical propagation of wavefronts near the umbra center. As the height augments, there occurs the oscillation source broadening, and the emergence of a quasi-spiral wave propagation. The cut-off frequency dependence on the magnetic field spatial distribution (polar angle) along which there occurs wave propagation, and on the emission generation height is assumed to be the main factor for the wavefront spiral-like form emergence.

\section{Sunspot waves and flare energy release}

The first ideas of flare initialization in active regions were described in \citep{1978A&A....68..145N}. Therein, there was argumentation that the energy release process cannot start throughout the entire flare volume simultaneously. At first, the flare is localized in a small plasma volume of an active region. Then, the energy release with its dissipation starts to capture new volumes of an active region. At least, two types of agents may lead to such a dissipative process: electron beams and shockwaves. These agents may result in the flare emergence at a great distance from the initial position of their source, causing simultaneous (sympathetic) flares in different active regions \citep{2009ApJ...702.1553L, 2009A&A...493..629Z}.

	For the first time, these trigger processes were numerically modeled in \citep{1989SoPh..124..319K, 1997A&A...326.1252O}. 
\citep{1989SoPh..124..319K} assumed that electron beams, passing through the current layers in the magnetic reconnection region, generate Langmuir waves. By using a numerical model, the authors studied the effect of these electrostatic waves on the plasma system.  Sufficiently strong Langmuir waves were found to be able to generate ion-sound waves \citep{2000A&A...353..757B}. These waves increase the electrical resistance in the current system, which results in the dissipative process start. Thereby, electron beams may cause a forced magnetic reconnection. 

\citep{2014ApJ...784..144D} showed the effect of the magnetic loops' periodic motions on the emergence of the so-called sliding reconnection between loops.	Analyzed were the microwave emission damping oscillations of a flare with a period about 12.5 minutes. Similar pulsations were observed in EUV in the SDO/AIA 335\r{AA} high-temperature channel \citep{2012ApJ...756L..36K}. The hot loops observed with SOHO/SUMMER were shown to be the oscillation source in the form of periodic Doppler shifts. For the first time, an association between slow magnetoacoustic waves in the flare microwave emission and the hot loop oscillations caused by standing waves was shown.

  The latest observation shown that such waves take place not only in hot and flaring coronal loops, but they are well exist in the comparatively cooler loops also.  The authors \citep{2010NewA...15....8S} using temporal image data from EIS/Hinode, for the first time detect the observational signature of multiple (first and second) harmonics of slow acoustic oscillations in the non-flaring coronal loop. The investigation suggests the simultaneous co-existence of standing oscillations and propagating bright blobs. Using 1-D loop hydrodynamic model and EIS/Hinode synthetic spectra, the observing possibilities of such oscillations were explored \citep{2008A&A...481..247T}
with prediction that slow acoustic oscillations should be observed with the imaging as well as spectral analysis of real Hinode data.

\citep{1997A&A...326.1252O} studied the flare initialization by shockwaves. The authors used a 2D magnetohydrodynamic model with shockwaves propagating through the current sheet. One part of the waves passed through the layer, the other reflected. Found was the emergence of wave-caused plasma flows near the current sheet. Those flows led to the magnetic reconnection emergence. One may conclude that a relevant reason leading to the reconnection emergence in plasma is not only an increase in the electric resistance, but also plasma flow emergence. 
	
	The 2005 August 22 long-lived flare observed in the radio-frequency range (NoRH, 17 GHz) and in X-ray (Ramaty High Energy Solar Spectroscopic Imager at 25-50 KeV) was studied in \citep{2011A&A...525A.112R}. An increase in the signal oscillation period from 2.5 to 5 minutes during the flare was found. Analyzing the quasi-periodic variations in the loop length and in the plasma temperature during the flare allowed the authors to interpret them as an emergence of the second harmonic of slow magnetoacoustic waves. This mode may be interpreted as a loop response to the initial impulse. The mode emergence served a trigger for periodic energy release.
	
	In \citep{2009ApJ...702....1H, 2011ApJ...743..142H, 2015MNRAS.446.3741C} the increasing 3-min wave oscillations from sunspots before the triggering of the jets was studied. This effect provide the evidence of wave induced magnetic reconnection. Observed dynamics can be interpreted as originating and/or contraction of new flux tubes, which location near  the neighbouring magnetic field lines may induce reconnection and formation of jets.
	
	An association between periodic initialization of reconnection and slow magnetoacoustic waves was revealed in \citep{2006SoPh..238..313C}. In this case, plasma density periodic disturbances in the reconnection region are observed. A similar correlation between the 3-min oscillations in the sunspot atmosphere, interpreted as slow magnetoacoustic waves and the flare short-period 3-min oscillations has been recently established \citep{2009A&A...505..791S}. In this work the first time, a phenomenological coupling between sunspot oscillations and quasi-periodic pulsations (QPP) of the flare energy release in the active region over the sunspot was shown. Analyzing the microwave emission recorded with the Nobeyama Radioheliograph at 17 GHz showed a significant increase in the sunspot 3-min oscillation train power in the active region prior to the flare. The flare curve also showed a modulation of the radio flux with a 3-min period. A spatial analysis of the 3-min oscillation distribution showed the formation of new V-shaped sources associated with the foot points of coronal loops in the 171\r{AA} line (SDO/AIA). Visible oscillations in the microwave range are propagating waves along coronal loops anchored in a sunspot toward the flare source. Sunspot oscillations were suggested to appear as a trigger of the flare radio emission. Due to the curvature of the magnetic field lines, emerging centrifugal forces may result in the formation of transverse oscillations due to the wave motion \citep{1989PAZh...15..154Z}. These oscillations, in turn, appear as a flare trigger due to reconnection emergence near the local magnetic zero-points \citep{2006A&A...452..343N}. Using this mechanism allowed one to explain  both the emergence of individual short bursts (single reconnections), and periodic 3-min modulation (forced cascade reconnections due to the wave energy supply) during flare evolution.

\begin{figure*}[t]
\label{fig6}
\epsfysize=5.5cm \centerline{\epsfbox{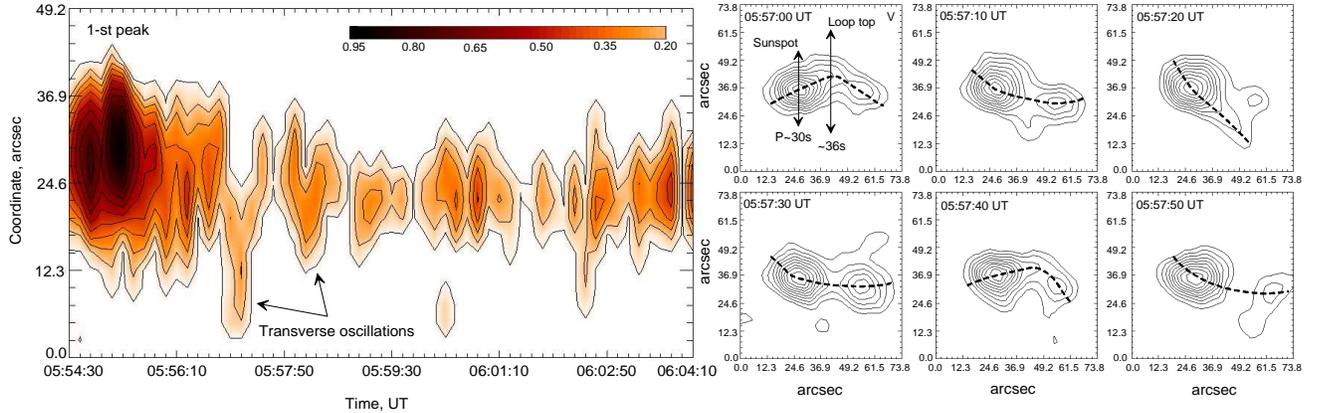}}
\caption{Flare loop forced transverse oscillations in the microwave range (NoRH, 17 GHz). {\it Left panel:} Space-time plot obtained by scanning the flare loop radio brightness cross section. {\it Right panel:} Variations in the radio loop two-dimensional structure over transverse oscillations (05:57:00-05:57:50 UT).{\bf \citep{2014arXiv1409.2947S}}}
\end{figure*}

	A statistical analysis of an increase in the 3-min sunspot oscillations before solar flare was performed in \citep{2011SoPh..273..403A}. The parameters of radio emission oscillations (NoRH, 17 GHz) in active regions, and their effect on the flare activity were studied. Illustrated by a large number of events, an increase in the 3-min oscillation power is shown to be characteristic of many small flares, which confirm the conclusions in \citep{2009A&A...505..791S}. A difference in the sunspot oscillations before the flare and after the main peak of the energy release was revealed. A correlation between the variations in the oscillation spectrum and the probability of flare emergence was found. The obtained results were interpreted within the model for the effect of the waves propagating from a sunspot on the flare region.
	
	A further endeavor to investigate the coupling of sunspot waves and flares was the study by \citep{2014arXiv1409.2947S}, where the authors performed a detailed spatial analysis of the magnetic waveguide emergence and evolution of propagating waves. The 3-min pulsations in the radio-frequency range (NoRH, 17 GHz) with the train period of $\sim$12-20 minutes were shown to have well-pronounced drifts of the oscillation period within $\sim$2-4 minutes. The drift value decreases before the flare. This indicates at the formation of the spatially allocated magnetic rope (channel), along which slow magnetoacoustic waves begin to propagate. The radio pulse source represents a loopy source, whose one foot point is anchored in the umbra. The waves propagate from the sunspot toward the flare region. For the first time, revealed were observational data on the transformation of longitudinal low-frequency 3-min waves propagating from a sunspot into transverse high-frequency loop oscillations with the $\sim$30-40 sec period (Fig. 6). In the coronal region (SDO/AIA, 171\r{AA}), the source of 3-min EUV oscillations coincides with a radio loop, and forms a fine-structure rope of magnetic loops (waveguides) connecting the sunspot to the flare region. A possible heating mechanism for the flare loop is shown to be the coupling (reconnections) of the initiated upper-tube transverse oscillations with the underlying twisted loops. As the reconnection region expands, the magnetic field configuration becomes simpler with the primary loop twist disappearance. A flat stage was revealed on the flux profile in soft X-ray after the main peak of flare emission. Its emergence is assumed to relate to maintaining additional energy release due to the mechanism for the forced quasi-periodic reconnections with the underlying loops, initiated by the flare loop transverse oscillations.
	
\subsection{Summary}

      There is a relation between the increase in the sunspot wave activity and the emergence of flares. Revealed was the existence of the sunspot atmosphere wave activity different phases. At the quiet phase, there occurs a continuous propagation of slow magnetoacoustic waves along the magnetic flux tubes (often with an open configuration) from the subphotospheric layers into the corona. The oscillation sources coincide with the foot points coronal loops anchored in the umbra. For the active phase, characteristic is a significant increase in the oscillation power within the microwave and an ultraviolet radiation with ~2-4 min periods. There was found the emergence of peak oscillations (by power) ~20-30 min prior to the beginning of the flare energy release. For this phase, characteristic is the emergence of V-shaped oscillation sources in the umbra. The wave propagation direction coincides with the flare source position. The slow low-frequency MHD-waves propagating from sunspots were revealed to be able to transform into transversal high-frequency oscillations, and to initiate the start of a forced reconnection in magnetic structures. The waves are shown to possibly be both the flare trigger and the emission modulator as the intensified wave flux reaches the flare source region via the magnetic waveguide. Also revealed was additional plasma heating with a production of flat patches at the stage of the X-ray flare decay due to continuation of the reconnection caused by the sunspot wave activity.
		
\section{Model of sunspot oscillations}

Sunspot oscillations were discovered by \citep{1969SoPh....7..351B}. Unlike 5-min oscillations in the quiet atmosphere, the sunspot oscillation period appeared to be equal to 3 minutes. However, soon it became clear that the oscillations, both in the quiet photosphere and in sunspots, are not only one-frequency oscillations. The spectrum of these oscillations comprises many spectral lines. Further, the study of the 5-min oscillation spectrum led to emergence of a new subdiscipline in solar physics, helioseismology. This occurred due to the creation of the oscillation p-mode theory and fast advance in the observational methods. Helioseismology and astroseismology, like classical spectroscopy of atoms and molecules, allows one to determine many properties of the Sun and stars use the oscillation spectrum.
	
	An important property of sunspot oscillations is that they are not global collective oscillations of the entire sunspot as a whole. Signatures of this were found earlier, but present-day observations at a high resolution do not leave any doubts that the sunspot oscillations are local, i.e. they represent an assemblage of many oscillating elements. Their spectra differ from one another. The sunspot oscillation theory should explain, first of all, the spectral composition of the oscillations and their locality. Currently, sunspot seismology is referred to as not classical seismology that allows one to determine the sunspot atmosphere structure by its spectrum, but rather studying the sunspot subphotospheric layers through local helioseismology with the oscillation p-modes. Sunspot local helioseismology is aimed at creating a model for the sunspot subphotospheric layers \citep{2010SoPh..267....1M}. In principle, the combination of the local seismology methods with the sunspot oscillation spectrum analysis should help investigate the sunspot structure.
	
	Unfortunately, there has been no sunspot oscillation model so far that explain  the entire oscillation spectrum, and not just 3-min oscillations. Within the chromospheric resonator model proposed by \citep{1981PAZh....7...44Z}, there have been endeavors to interpret the 3-min oscillation spectrum \citep{2001SoPh..202..281S, 2007AstL...33..622Z, 2008SoPh..251..501Z}. In this theory, the disturbances of the medium caused by irregular plasma motions in the subphotospheric layers may generate waves that permeate into the upper atmosphere only at certain frequencies. The sunspot chromosphere is envisioned as the Fabry-Perrot filter for propagating slow magnetoacoustic waves in strong magnetic fields \citep{1983SoPh...82..369Z}. Recent numerical calculations \citep{2011ApJ...728...84B} showed that different profiles of plasma temperature and density in the wave propagation medium result in different oscillation frequency and efficiency of the wave permeation into the corona. Therefore, the sunspot atmosphere horizontal inhomogeneity leads to emerging a fine spectral structure of oscillations, and the oscillation power distribution over a sunspot \citep{2008SoPh..251..501Z, 2014A&A...569A..72S}. Besides, this structuring may be different at different heights.
	
	In order to understand the influence of magnetic fields on the propagation properties of waves forward modeling of waves is required. 
Such calculations need a model in magnetohydrostatic equilibriumas an initial atmosphere through which to propagate oscillations.
In \citep{2008ApJ...689.1379K} developed a method to construct a model in equilibrium for a wide range of physical parameters. The method combines the advantages of self-similar solutions and current-distributed models. A set of models is developed by numerical integration of magnetohydrostatic equations from the subphotospheric to chromospheric layers.

   Further studies \citep{2008ApJ...676L..85K} with use two-dimensional MHD simulations have shown that photospheric 5 minute oscillations can leak into the chromosphere inside small-scale vertical magnetic flux tubes. The results of numerical experiments are compatible with those inferred from simultaneous spectropolarimetric observations of the photosphere and chromosphere obtained with the Tenerife Infrared Polarimeter (TIP) at 10830\r{AA}. The efficiency of energy exchange by radiation in the solar photosphere can lead to a significant reduction of the cutoff frequency with the propagation of the 5 minute waves vertically into the chromosphere.
		
	The urgency of sunspot chromosphere seismology is related to that building an empirical model of the chromosphere through classical methods challenges serious difficulties: the models developed by different authors \citep{1986ApJ...306..284M, 1981A&A...100..284S} substantially differ from one another \citep{2007AstL...33..622Z, 2008SoPh..251..501Z}. However, chromosphere seismology within the chromospheric resonator model also faced problems, because it failed to explain all the complex spectrum of sunspot oscillations. Moreover, these studies were based on observations of sunspot oscillations \citep{2006RSPTA.364..313B} at a relatively low spectral resolution, which, as it appeared later, did not allow one to obtain a detailed oscillation spectrum. Only after the SDO/AIA launch, it became possible to obtain spectra at a significantly larger spectral resolution. The 3-min oscillation spectrum appeared to involve much more spectral harmonics, than it had been considered earlier \citep{2012ApJ...746..119R, 2014A&A...561A..19Y}.
	
	\citep{2014AstL...40..576Z} performed a spectral analysis of the data with a high temporal and spatial resolution of the ultraviolet oscillations at the transition level (SDO/AIA, 304\r{AA}). The obtained spectrum of 3-min oscillations involves a large number of narrowband harmonics. There is no detected single 3-min peak. The oscillations concentrate in $\sim$1500 km spatial cells with almost identical frequencies, without global oscillations of the entire sunspot. Oscillation cells emerge in those locations, where advancing plasma, moving apart the magnetic field, reaches the sunspot photosphere lower boundary. In these plasma regions, there appears an increased temperature gradient and a weakened magnetic field. There exists a set of hot plasma fluxes with a weak magnetic field surrounded by a relatively strong magnetic field. It is possible that the oscillation cells are related to bright umbral dots.
	
	The chromospheric resonator is shown not to be responsible for the 3-min oscillation spectrum, because it cannot provide such a number of spectral harmonics within the 1.7-3.5 min period band. Numerical calculations \citep{2001SoPh..202..281S} show that the chromospheric resonator leads to the emergence of only several spectral harmonics in this frequency band, but not tens of harmonics like it occurs in reality. The existence of subphotospheric resonators may result in the emergence of the spectrum with a large number of spectral lines. This is possible in case of a quite depth-extended resonator with a sufficiently low main resonance frequency. In this case, numerous spectral harmonics may be a result of excitation high harmonics in the low-frequency subphotospheric resonator representing a system of two coupled resonators. One may expect the existence of association between 3-minute and low-frequency oscillations that are also observed in sunspots \citep{2013PASJ...65S..13B}.
	
	The obtained diversities of oscillation spectra in different parts of a sunspot find a natural explanation within the Parker sunspot model, according to which the sunspot magnetic field is split into individual magnetic tubes in the subphotospheric layers. These magnetic tubes cannot be identical. The differences between the thin tubes should concern the magnetic field value, temperature gradients, and density. These parameters determine the properties of the subphotospheric resonator, and the cut-off frequency and phase velocity of slow magnetoacoustic waves depend on them. In other words, the pattern of sunspot oscillation distribution is the reflection of the magnetic field structure above a sunspot.
	
\subsection{Summary}

The investigations into the 3-min sunspot oscillations with a high spatial and temporal resolution (SDO/AIA) showed that the model of the chromospheric resonator \citep{1981PAZh....7...44Z} is not able to elucidate the observations. Studying the oscillation spectrum does not corroborate the existence of the detected peak near the 3-min period. There is a set of narrowband spectral harmonics within the 2-4 minute range. The sunspot has no global oscillations as a whole. For each harmonic, one can find a correspondence to a small angle-size umbra site oscillating at the given frequency. To elucidate the locality of 3-min oscillations, there is a suggestion of a slow-wave resonance that originates in the subphotospheric magnetic tubes (the Parker model) and encompasses all the sunspot atmosphere layers. In this case, the observed numerous spectral lines may be a result of high harmonics excitation in the low-frequency subphotospheric resonator.

\begin{acknowledgments}
This work was supported by the Russian Foundation for Basic Research under Grants 13-02-00044, 13-02-90472, 14-02-91157. The research was funded by Chinese Academy of Sciences President's International Fellowship Initiative (PIFI) grant for Visiting Scientists, grant No. 2015VMA014.
\end{acknowledgments}



\end{article}
\end {document}